\documentclass{article}
\usepackage{spconf,amsmath,graphicx, booktabs}

\title{A Comprehensive Study of Content Representations for Speech Synthesis}
\name{Diego Torres  \qquad Axel Roebel \qquad Nicolas Obin}
\address{STMS Lab \\
      IRCAM, CNRS, Sorbonne Université \\ Paris, France}
\begin{document}
\ninept
\maketitle
\begin{abstract}
Speech content representations are central to voice conversion, speech-to-speech translation, and multimodal language models, yet they are rarely compared under a common generative framework that directly measures what each representation contains. We address this by training a generative model conditioned solely on each representation and evaluating the generated audio along the content, speaker identity, and prosody axes. Across SSL features, supervised tokens, posteriorgrams, and neural audio codecs, we find two distinct regimes: representations that nearly reconstruct the original audio, and representations that effectively disentangle speaker identity. These results show that disentanglement depends not on supervision alone, but on the interaction between the training objective and the representation's information capacity: supervised representations only disentangle speaker identity when their capacity is sufficiently constrained.
\end{abstract}
\begin{keywords}
speech synthesis, speech tokenizer, voice conversion, speaker disentanglement
\end{keywords}

\vspace{-6pt}
\section{Introduction}
\label{sec:intro}

Content representations of speech (often called ``semantic tokens''\footnote{We prefer ``content representations'' as these features encode phonetic rather than semantic information \cite{phonetic_not_semantic}, and are not necessarily discrete.}) aim to capture the linguistic content of a speech signal while discarding speaker, prosodic, and acoustic information. They have facilitated speech synthesis tasks such as voice conversion \cite{sun2016phonetic, qian2022contentvec}, style transfer \cite{vevo}, and speech-to-speech translation \cite{speech2speech_trans}, and have also been used as intermediate representations in text-to-speech systems \cite{speartts, du2024cosyvoice}. Moreover, combined with LLMs, they have enabled multimodal models capable of performing a wide range of speech- and sound-related tasks \cite{borsos2023audiolm, zhang2023speechgpt, rubenstein2023audiopalm}.
Different applications, however, impose different and sometimes opposing requirements: conventional voice conversion focuses primarily on timbral information, with prosodic changes generally being incidental, whereas accent conversion explicitly targets prosodic characteristics. The relevant question is therefore not whether a representation is disentangled in some abstract sense, but rather which aspects of speech it captures, and to what extent.

Prior work has examined this question from several angles. One line of research has benchmarked a wide range of speech tokenizers on discriminative and generative tasks \cite{mousavi_1, mousavi_dasb, mousavi_survey}. However, these studies place acoustic and content tokenizers in a single ranking despite their differing objectives, and do not directly assess disentanglement. Another study \cite{estimating} probes SSL features using prediction heads, but is limited to k-means and RVQ representations derived from two HuBERT layers, leaving a broad range of representations unexplored. Similarly, \cite{rethinking} measures the recoverability of accent information in synthesized speech using ABX testing and a generative framework. Beyond these benchmarking efforts, Soft-VC \cite{softvc} and Vevo \cite{vevo} investigate how architectural choices affect disentanglement in the context of voice conversion: Soft-VC compares discrete and soft representations, while Vevo examines the effect of codebook size for discrete representations.

This paper builds on and consolidates these efforts. Existing evaluations largely group representations according to how they are produced, focusing primarily on SSL features and their quantized variants. In contrast, we adopt a broader scope, considering a diverse range of representations, including clustering-based representations, supervised tokens, posteriorgrams, and neural audio codecs. Although these representations arise from fundamentally different approaches, they can be viewed as alternative ways of encoding linguistic content for downstream applications. Despite this shared role, they have rarely been evaluated under a common speech synthesis framework. Our unified evaluation reveals distinct regimes among these representations, shaped primarily by the interplay between training objective and information capacity.

\vspace{-5pt}
\section{Methodology}
\vspace{-5pt}

\begin{figure}
    \centering
    \includegraphics[width=0.7\linewidth]{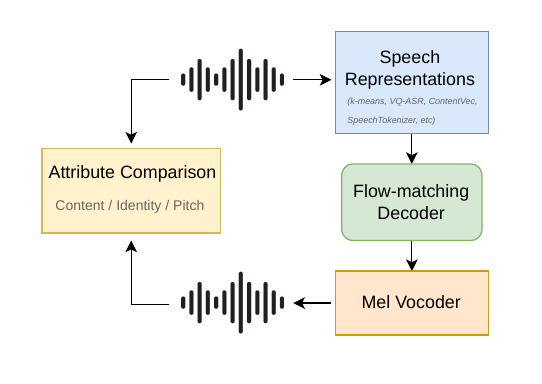}
    \caption{Diagram of the evaluation methodology.}
    \label{fig:methodology}
    \vspace{-15pt}
\end{figure}

Our methodology builds on the approach used by \cite{softvc, vevo, rethinking}, in which a generative model is trained to synthesize speech from a given representation, and the generated audio is subsequently evaluated for the attributes it preserves (Figure~\ref{fig:methodology}). Unlike these approaches, our generative model has no access to speaker identity information, either in the form of a speaker ID (as in \cite{rethinking}) or reference audio (as in \cite{vevo}). Providing such information introduces a potential confound: even if the representation contains speaker-specific information, the generative model may learn to rely on the additional identity channel as well. In that case, evaluating voice conversion performance does not provide a reliable estimate of the degree of speaker disentanglement in the representation itself.\footnote{This does not invalidate their results for voice conversion, but rather limits their interpretation as a measure of representation-level disentanglement.} 
Our setup avoids this confound, as any speaker information present in the generated audio must necessarily pass through the representation.

Formally, the generative model learns the conditional distribution $x_\text{gen} \sim p(x|r)$ given the representation $r$. Since $r$ is a deterministic function of the input audio, $r=f(x)$, generation can be viewed as sampling from the set of utterances compatible with it, $\mathcal{X}(r)=\{x:f(x)=r\}$. Attributes that are invariant within this set are preserved in the generated sample, whereas those that vary across the set are resampled according to the learned distribution. 

For example, if the representation encodes speaker identity, the generated speech is expected to preserve the identity of the source speaker, while other attributes, such as prosody, may be resampled. Conversely, if speaker identity is only partially encoded, the generated audio may not reproduce the original speaker exactly, but it may share coarse attributes with the source, such as gender, accent, or vocal characteristics, without fully matching the source identity \footnote{Audio samples: https://content-demo.diegotg2000.workers.dev/}.
In practice, the expression of an attribute in generated audio depends not only on the representation but also on the decoder’s capacity to recover it, which makes the evaluation a lower bound on the true amount of information inside the representations.

\vspace{-10pt}
\subsection{Evaluation metrics}

We compare the original and the generated audio along three axes: linguistic content, speaker identity and prosody. Content is measured by the differential word error rate (dWER), which is the word error rate between the ASR transcriptions of the two audios. Evaluation is restricted to the utterances that the ASR transcribes correctly from the real audio. We do this to prevent audios of poor quality skewing the distribution in favor of acoustic features that simply recreate the audio, even when content is gibberish. Speaker identity is measured with the VoicePrivacy protocol \cite{voiceprivacy}: the equal error rate (EER) of a speaker verification system asked to verify the generated audio against the original speaker, scoring each trial by the cosine similarity between the two embeddings. Target trials pair a generated utterance with a real utterance of the same speaker, non-target trials with a real utterance of a different speaker. An EER of $50\%$ means the verifier is at chance and no speaker information was present in the input representation. We also fit a logistic probe to predict gender on the speaker embeddings and compute its accuracy against the gender of the source speaker. The evaluation dataset is almost gender-balanced, with 52.9\% of utterances being female, which gives the floor of the probe accuracy. As for prosody, we decompose the MSE error between the F0 contours (in cents) of the original and generated audio as:
{
\setlength{\abovedisplayskip}{5pt}
\setlength{\belowdisplayskip}{5pt}
\begin{equation*}
\text{MSE} = (\mu_\text{og} - \mu_\text{gen})^2 + (\sigma_\text{og} - \sigma_\text{gen})^2 + 2\sigma_\text{og}\sigma_\text{gen}(1-\rho),
\end{equation*}
}
where $\mu$ and $\sigma$ denote the mean and standard deviation of the F0, and $\rho$ denotes the correlation between the two contours. We refer to $\Delta\mu$ as the register error and $\Delta\sigma$ as the range error.

Transcriptions are produced by Parakeet\footnote{https://huggingface.co/nvidia/parakeet-ctc-0.6b} \cite{parakeet}, speaker embeddings by ECAPA-TDNN\footnote{https://huggingface.co/speechbrain/spkrec-ecapa-voxceleb} \cite{ecapa} and F0 by PENN\footnote{https://github.com/interactiveaudiolab/penn}\cite{penn}. We additionally report UTMOS\footnote{https://github.com/Blinorot/utmos-pytorch} \cite{utmos} as a check that generated audio remains of decent quality without severely affecting the other metrics.

\vspace{-8pt}
\section{Implementation details}

\vspace{-5pt}
\subsection{Data}

We perform our experiments on the LibriTTS dataset \cite{libritts}. Training is done on the train.clean.100 and train.clean.360 splits, model checkpoints are chosen by their loss on dev.clean, and the final evaluation is reported on test.clean. This includes training and validation of the generative model as well as the representations and vocoder. Audio is used at 24~kHz for synthesis and resampled to 16~kHz wherever a self-supervised encoder is involved.

\vspace{-5pt}
\subsection{Representations}

All representations we study operate at 50~Hz, allowing us to compare the information encoded per frame directly, and avoid the confounding of the frame rate.

\noindent\textbf{SSL features.}
We evaluate HuBERT-base~\cite{hsu2021hubert} and WavLM-base~\cite{chen2022wavlm} representations, using the sixth layer following related work~\cite{polyak}. We also apply k-means clustering to these features using the FAISS library~\cite{faiss}, with 10k randomly sampled utterances from train.clean.360 and 20 iterations. Since preliminary results showed no significant difference between the two models, we use HuBERT-base layer 6 for all subsequent experiments where an SSL model is used as backbone.

\noindent\textbf{VQ-VAE.}
Inspired by Vevo~\cite{vevo}, we train an autoencoder with a VQ bottleneck to reconstruct HuBERT features using an L2 loss. The architecture consists of eight 1D ConvNeXt blocks~\cite{vocos} with 512 channels, with the quantizer placed in the middle. The codebook has 256 dimensions and is updated using EMA, while the commitment loss is weighted by 0.25.

\noindent\textbf{VQ-ASR.}
Inspired by CosyVoice~\cite{du2024cosyvoice}, we train an ASR model with a VQ bottleneck. The model takes frozen HuBERT features as input and uses the same architecture as VQ-VAE, followed by a linear CTC head over characters. We additionally consider the continuous latent features \emph{before} the VQ module, which allows us to separately study supervision without quantization.

\noindent\textbf{PPG.}
We use MFA alignments~\cite{mfa} as targets to train a frame-wise phoneme recognizer on frozen HuBERT features. The model consists of eight ConvNeXt blocks with 512 channels, with a self-attention layer inserted every two blocks. Its output is a 40-dimensional posterior distribution per frame over the 39 ARPAbet phones plus silence. We also evaluate the argmax variant, which yields a discrete representation over 40 symbols

\noindent\textbf{Soft units.}
Introduced in Soft-VC~\cite{softvc}, we fine-tune HuBERT to predict $k$-means indices, enforcing a ``semantic'' yet continuous representation. The targets are obtained from the 64-cluster $k$-means model described above. Due to computational constraints the fine-tuning is done with LoRA~\cite{lora} (rank 16, $\alpha = 32$, dropout 0.05) on the attention projections of every layer. Further implementation details can be found in~\cite{softvc}.

\noindent\textbf{SpeechTokenizer.}
Introduced in~\cite{speechtokenizer}, SpeechTokenizer is a neural audio codec based on RVQ, with its first codebook level distilled from HuBERT and intended to capture linguistic content. We use the publicly available checkpoint\footnote{https://huggingface.co/OpenMOSS-Team/SpeechTokenizer} trained on LibriSpeech and retain only the first level.

\noindent\textbf{ContentVec.}
Introduced in \cite{qian2022contentvec}, its training recipe is specifically designed to derive speaker-independent content representations. We use the publicly available ``legacy'' checkpoint trained on LibriSpeech with 100 clusters as targets\footnote{https://github.com/auspicious3000/contentvec}.

\subsection{Acoustic model}

The same acoustic model is trained on all representations, only the input dimensionality changes. Content representations are first upsampled from 50~Hz to 100~Hz, after which a flow-matching module~\cite{flowmatching} uses them as conditioning to generate 100~Hz mel-spectrograms (100 mel bands, 1024-point FFT, hop size 240 at 24~kHz). Both components are built from ConvNeXt layers with an attention layer. The upsampler comprises two blocks of six layers, while the flow-matching module uses a U-Net architecture with four blocks of eight layers and $\sigma_{min}=10^{-4}$. The resulting model has approximately 67M parameters and is trained for 800k iterations with a batch size of 32 on a single RTX 4070, using AdamW with a learning rate of $10^{-4}$. We use classifier-free guidance~\cite{cfg}, dropping the conditioning with probability 0.1.

At inference, we solve the flow ODE using a midpoint solver with 10 steps and a guidance scale of $w=2.0$, using the same fixed random seed for all systems. Waveforms are generated with a custom Vocos~\cite{vocos} checkpoint trained for 1M steps.

\begin{table}[t]
\centering
\caption{Evaluation results for the considered speech content representations. dWER, EER and gender accuracy are measured in percentage. Register and range errors are measured in cents. dWER is computed corpus-wise, prosody metrics are shown as the median, while gender and speaker similarity are the mean. $K$ indicates the number of discrete codes, and is set to $\infty$ whenever a representation is continuous.
}
\vspace{5pt}
\label{tab:canonical-small}
\footnotesize
\setlength{\tabcolsep}{3pt}
\begin{tabular}{@{}llrrrrrrr@{}}
\toprule
 &  & Content & \multicolumn{3}{c}{Identity} & \multicolumn{3}{c}{Prosody} \\
\cmidrule(lr){3-3}\cmidrule(lr){4-6}\cmidrule(lr){7-9}
 & $K$ & dWER & EER & Gender & Sim. & Reg. & Range & Corr. \\
\midrule
Real audio & -- & -- & 0.5 & 96.9 & -- & -- & -- & -- \\
Vocos & -- & 0.2 & 0.6 & 97.8 & 0.89 & 6 & 8 & 0.98 \\
\addlinespace
HuBERT & $\infty$ & 0.5 & 1.3 & 98.7 & 0.61 & 48 & 30 & 0.94 \\
     & 1024 & 1.7 & 32.4 & 91.7 & 0.16 & 181 & 81 & 0.48 \\
     & 256 & 2.3 & 36.1 & 78.4 & 0.14 & 206 & 90 & 0.41 \\
     & 64 & 5.8 & 36.0 & 84.0 & 0.15 & 261 & 108 & 0.40 \\
\addlinespace
WavLM & $\infty$ & 0.5 & 2.0 & 99.0 & 0.52 & 71 & 29 & 0.94 \\
 & 256 & 1.9 & 37.4 & 79.8 & 0.14 & 263 & 93 & 0.41 \\
\addlinespace
VQ-ASR & $\infty$ & 0.7 & 8.6 & 99.0 & 0.35 & 100 & 54 & 0.83 \\
 & 1024 & 2.2 & 46.3 & 55.0 & 0.11 & 450 & 100 & 0.36 \\
 & 256 & 2.7 & 46.6 & 53.0 & 0.11 & 452 & 99 & 0.33 \\
 & 64 & 3.0 & 48.4 & 49.5 & 0.10 & 505 & 96 & 0.34 \\
\addlinespace
VQ-VAE & 1024 & 0.8 & 12.8 & 99.3 & 0.30 & 117 & 58 & 0.88 \\
 & 256 & 1.1 & 15.7 & 99.2 & 0.27 & 127 & 59 & 0.87 \\
 & 64 & 1.4 & 20.3 & 98.9 & 0.24 & 136 & 71 & 0.84 \\
\addlinespace
Soft units & $\infty$ & 0.5 & 2.2 & 98.5 & 0.51 & 79 & 42 & 0.92 \\
\addlinespace
PPG & $\infty$ & 2.0 & 40.0 & 74.9 & 0.13 & 328 & 90 & 0.31 \\
\quad argmax & 40 & 2.9 & 43.0 & 68.6 & 0.12 & 455 & 183 & 0.26 \\
\addlinespace
ContentVec & $\infty$ & 0.6 & 16.3 & 91.0 & 0.28 & 177 & 75 & 0.88 \\
\addlinespace
SpeechTok. & 1024 & 3.1 & 28.8 & 92.7 & 0.18 & 204 & 89 & 0.41 \\
\bottomrule
\end{tabular}
\vspace{-10pt}
\end{table}

\begin{figure}
    \centering
    \includegraphics[width=0.95\linewidth]{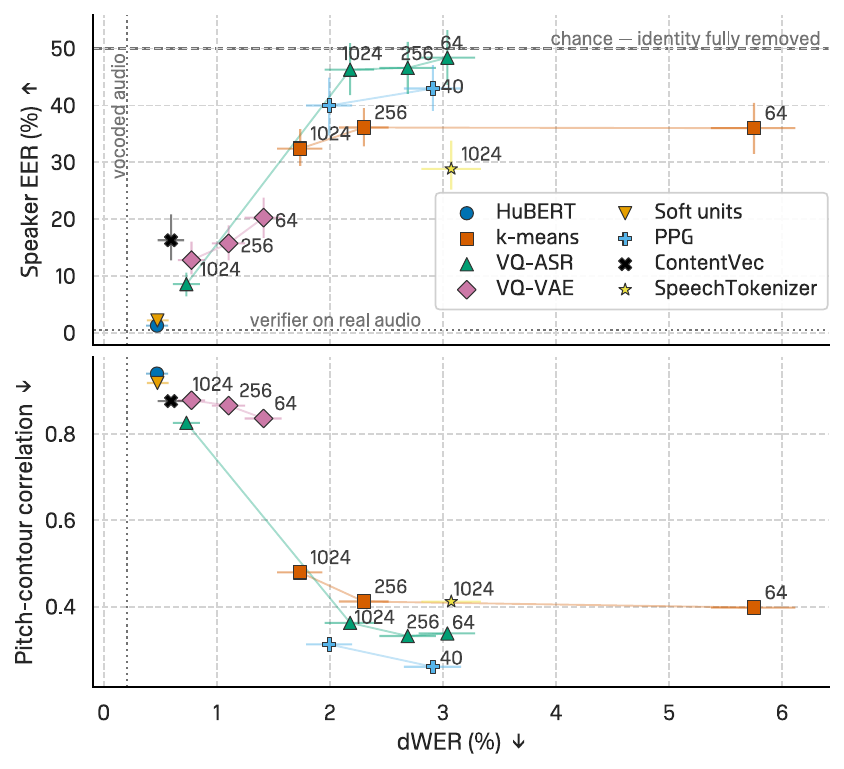}
    \caption{Scatter plot of EER and pitch correlation versus dWER. Numbers next to the markers indicate the codebook size, and their absence indicates that the features are continuous. Bars indicate the 95\% confidence intervals of: the mean of EER, the median of pitch correlation, and the corpus-wise dWER.}
    \label{fig:tradeoffs}
    \vspace{-15pt}
\end{figure}


\begin{table}[t]
\centering
\caption{Evaluation of VQ-ASR when the codebook dimension is decreased to 8. Both the quantized version with 256 codes and the continuous version are considered.}
\vspace{3pt}
\label{tab:vq-d8}
\footnotesize
\setlength{\tabcolsep}{5pt}
\begin{tabular}{@{}llrrrr@{}}
\toprule
 & $K$ & dWER (\%) & EER (\%) & Gender (\%) & Pitch Corr. \\
\midrule
VQ-ASR 8-d & 256 & $\uparrow$\,\textbf{3.2} & 46.8 & 55.9 & $\downarrow$\,\textbf{0.29} \\
 & $\infty$ & $\uparrow$\,\textbf{2.4} & $\uparrow$\,\textbf{39.7} & $\downarrow$\,\textbf{71.9} & $\downarrow$\,\textbf{0.35} \\
\bottomrule
\end{tabular}\\[4pt]
{\scriptsize \textbf{Bold}: statistically significant differences with respect to the equivalent 256-$d$ VQ.}
\vspace{-12pt}
\end{table}

\begin{table}[t]
\centering
\caption{Comparison of metrics when guidance scale is changed ($w=1.0$ vs $w=2.0$).}
\vspace{5pt}
\label{tab:guidance}
\footnotesize
\setlength{\tabcolsep}{3pt}
\begin{tabular}{@{}lrrrrrrr@{}}
\toprule
 &  & \multicolumn{2}{c}{dWER (\%)} & \multicolumn{2}{c}{EER (\%)} & \multicolumn{2}{c}{UTMOS} \\
\cmidrule(lr){3-4}\cmidrule(lr){5-6}\cmidrule(lr){7-8}
Representation & bit/s & $w{=}1$ & $w{=}2$ & $w{=}1$ & $w{=}2$ & $w{=}1$ & $w{=}2$ \\
\midrule
Real audio & -- & \multicolumn{2}{c}{--} & \multicolumn{2}{c}{0.5} & \multicolumn{2}{c}{4.14} \\
Vocos & -- & \multicolumn{2}{c}{0.2} & \multicolumn{2}{c}{0.6} & \multicolumn{2}{c}{3.77} \\
\midrule
HuBERT & 1.2\,M & 0.5 & 0.5 & 1.3 & 1.3 & 3.77 & 3.77 \\
ContentVec & 1.2\,M & 1.0 & $\downarrow$\,\textbf{0.6} & 19.6 & $\downarrow$\,\textbf{16.3} & 3.54 & $\uparrow$\,\textbf{3.86} \\
Soft units & 410\,k & 0.5 & 0.5 & 3.2 & $\downarrow$\, \,\textbf{2.2} & 3.74 & $\uparrow$\,\textbf{3.82} \\
PPG & 64\,k & 3.4 & $\downarrow$\,\textbf{2.0} & 40.5 & 40.0 & 3.42 & $\uparrow$\,\textbf{4.00} \\
k-means 256 & 400 & 3.1 & $\downarrow$\,\textbf{2.3} & 38.7 & 36.1 & 3.48 & $\uparrow$\,\textbf{4.02} \\
VQ-VAE 256 & 400 & 1.6 & $\downarrow$\,\textbf{1.1} & 18.8 & $\downarrow$\,\textbf{15.7} & 3.54 & $\uparrow$\,\textbf{3.94} \\
VQ-ASR 256 & 400 & 4.2 & $\downarrow$\,\textbf{2.7} & 46.2 & 46.6 & 3.24 & $\uparrow$\,\textbf{4.04} \\
\bottomrule
\end{tabular}\\[4pt]
{\scriptsize \textbf{Bold}: statistically significant differences between $w{=}1.0$ and $w{=}2.0$.}
\vspace{-12pt}
\end{table}

\begin{table}[t]
\centering
\caption{Evaluation with a decoder with three times as many parameters.}
\vspace{5pt}
\label{tab:capacity}
\footnotesize
\setlength{\tabcolsep}{5pt}
\begin{tabular}{@{}llrrrr@{}}
\toprule
 & $K$ & dWER (\%) & EER (\%) & Gender (\%) & Pitch Corr. \\
\midrule
$k$-means & 256 & 2.3 & 34.8 & 85.0 & $\uparrow$\,\textbf{0.43} \\
VQ-ASR & 256 & $\downarrow$\,\textbf{2.5} & 47.7 & 50.5 & $\uparrow$\,\textbf{0.37} \\
\bottomrule
\end{tabular}\\[4pt]
{\scriptsize \textbf{Bold}: statistically significant differences with respect to the normal decoder. }
\vspace{-12pt}
\end{table}

\vspace{-5pt}
\section{Results}
\vspace{-5pt}

Our results reveal that the considered representations span the full range of disentanglement, from features that nearly preserve the original audio to those that effectively mask speaker identity. Raw HuBERT and WavLM features fall in the first group, which is expected given that neural audio codecs \cite{focalcodec} and waveform vocoders \cite{knnvc} have been trained to reconstruct audio from such representations, confirming that they retain sufficient information for high-fidelity reconstruction. Surprisingly, Soft~Units lie close to these representations, while ContentVec and continuous VQ-ASR also remain closer to this group than to the more disentangled representations. These representations share a continuous, high-dimensional structure, suggesting that supervision promoting linguistic content does not by itself prevent other acoustic information from being encoded. At the other extreme, VQ-ASR and PPGs effectively remove speaker information from the generated audio: VQ-ASR achieves EER values indistinguishable from chance, while its gender-classification accuracy remains close to the 52.9\% baseline determined by the gender distribution. This suggests that it is the \emph{combination} of supervision and a bottleneck that drives disentanglement \cite{pitchflower}. Crucially, a bottleneck alone is not sufficient either. Although VQ-VAE imposes a bottleneck, its sole objective is feature reconstruction, resulting in effective compression \cite{focalcodec} but little disentanglement.

Codebook size has a secondary effect: smaller codebooks tend to reduce speaker and pitch information, but the effect is insufficient to move a representation between the two regimes. This contrasts with approaches such as Vevo \cite{vevo}, which use codebook size to differentiate content from content-style tokens. Within their task's framework, our results suggest a natural assignment: VQ-ASR features serve as content tokens, whereas VQ-VAE features are better categorized as content-style tokens. 


We also investigate the effect of reducing the dimensionality of the VQ module, from 256 to only 8 dimensions. We evaluate both the quantized and continuous variants, with the results shown in Table~\ref{tab:vq-d8}. The performance of the quantized version is largely unaffected by this substantial reduction in dimensionality, with dWER and pitch correlation being the only metrics showing significant differences. In contrast, the continuous version exhibits a markedly different behavior: the 8-dimensional variant achieves substantially better disentanglement performance than its 256-dimensional counterpart, with EER going from 8.6\% to 39.7\%. Together with the performance of continuous PPGs, these results support the hypothesis that the main contribution of quantization is to constrain representational capacity, rather than to directly induce disentanglement. 

Interestingly, the first RVQ level of SpeechTokenizer falls outside of the two groups. At 1024 codes, VQ-VAE provides substantially better performance as a ``complete'' representation, whereas VQ-ASR and even $k$-means perform better as linguistic representations, achieving lower dWER at comparable or higher EER. One possible explanation is that SpeechTokenizer's first level is jointly optimized for signal reconstruction and HuBERT distillation. These objectives encourage the representation to retain both acoustic and linguistic information, resulting in a code that is neither strongly phonetic nor fully acoustic. 

To understand why VQ-ASR and PPGs, which are explicitly trained to capture linguistic content, exhibit higher dWER, qualitative inspection suggests that the main source is encoder-level confusion. In fast or out-of-distribution speech, frames can be assigned to incorrect linguistic clusters, causing the generator to reconstruct audio that is plausible but linguistically incorrect. This failure mode is largely absent for the ``complete'' representations, which do not attempt to extract discrete linguistic units and instead preserve sufficient information for near-complete reconstruction. Regarding prosody, as measured by pitch correlation, we find that zero correlation is not attainable: linguistic content inherently constrains the overall shape of the pitch contour. The correlation observed for PPGs can therefore be interpreted as a practical lower bound on the amount of pitch information, which VQ-ASR approaches closely.

\vspace{-5pt}
\subsection{Robustness to decoder capacity and sampling parameters}

Having established the main findings, we now verify their robustness to decoder capacity and sampling parameters. First, we train a decoder with three times as many parameters ($\sim$221M) on top of the $k$-means and VQ-ASR features with 256 codes. As shown in Table~\ref{tab:capacity}, most of the metrics show no significant change, and even where differences are statistically significant, they remain small compared to the gaps between the representations. 

As for sampling parameters, we run inference at guidance scales of $w = 1.0$ and $w = 2.0$ (Table~\ref{tab:guidance}). Increasing guidance generally improves both dWER and UTMOS across all representations, but does not change their relative ordering: the only pair that swaps, PPGs and $k$-means 256, was statistically indistinguishable at $w = 1.0$. EER is mostly unaffected, except for a significant yet small decrease for the ``complete'' representations. Interestingly, at $w = 2.0$, disentangled representations achieve a UTMOS score even higher than ground-truth Vocos reconstruction. This is likely because a more ``complete'' representation nearly determines the recording, leaving guidance little room for improvement, whereas a bottlenecked representation leaves some details unspecified. Guidance can then steer these details towards a clean, typical realization rather than reproducing the noise and room characteristics of the original recording.

\vspace{-10pt}
\section{Limitations}
\vspace{-5pt}

Our methodology provides a lower bound on the information present in each representation: only a perfect decoder and a perfect evaluator would allow us to measure the true information content. In practice, the decoder may fail to exploit information that the representation does contain, and the evaluation metrics are themselves imperfect proxies. Results with a larger decoder (Table~\ref{tab:capacity}) and varying guidance (Table~\ref{tab:guidance}) suggest that the gaps we observe are not driven by these factors, as they leave the relative ordering of representations largely unchanged, but the lower-bound interpretation should be kept in mind.

A key advantage of our generative framework is that it produces audio, which opens the door to perceptual evaluation. The objective metrics we rely on (UTMOS for quality, or ECAPA-TDNN for speaker identity) are proxies for human judgment, and may not capture all perceptually relevant distinctions. Subjective listening tests would provide a perceptual measure of what is preserved and discarded. We leave this extension to future work.

\vspace{-5pt}
\section{Conclusions}
\vspace{-5pt}

We presented a standardized, reference-free generative framework to benchmark speech content representations. Our findings offer several practical guidelines for speech tokenization:
~\textit{(1)~Supervision requires a bottleneck:} ASR objectives get the best disentanglement properties, but only when an explicit constraint, such as vector quantization or low-dimensional projection, is used to eliminate identity and prosodic leakage.
~\textit{(2)~Decouple compression from content:} Jointly optimizing single tokens for reconstruction and distillation (as in SpeechTokenizer's first layer) degrades both linguistic accuracy and disentanglement relative to dedicated content codes.
~\textit{(3)~Codebook size is secondary:} Varying codebook capacity changes the information content within a regime but does not alter the fundamental disentanglement profile set by the training objective.
Overall, this unified information space facilitates the choice of representations for downstream tasks and provides a standardized benchmark for evaluating future representations.

\clearpage

\section{Acknowledgments}
This work was partly performed using HPC resources from GENCI-IDRIS (Grant 2026-AD011016144R1), and funded by the ANR project EVA (ANR-23-CE23-0018).

{
\setlength{\itemsep}{0pt}
\bibliographystyle{IEEEtran}
\bibliography{refs}}

\end{document}